# Unique Sense: Smart Computing Prototype for Industry 4.0 Revolution with IOT and Bigdata Implementation Model

S. Vijaykumar[1]*, S. G. Saravanakumar[2] and M. Balamurugan[1]

[1]School of Computer Science, Engineering and Applications, Bharathidasan University, Trichy – 23, Tamil Nadu, India; indianid@gmail.com, mmbalmurugan@gmail.com
[2]6th SENSE, An Advanced Research and Scientific Experiment Foundation, Kumbakonam, Tamil Nadu, India; saravanakumarsg@gmail.com

## Abstract

Today, The Computing architectures are one of the most complex constrained developing area in the research field. Which delivers solution for different domains computation problem from its stack above. The architectural integration constrains makes difficulties to customize and modify the system for dynamic industrial and business needs. This model is the initiation towards the solution for findings of Industry 4.0 and Bigdata needs. This "Unique sense" smart computing implementation model for Industry 4.0 holds the innovative Smart computing prototype is a part of "UNIQUE SENSE" computing architecture which can delivers alternate solution for today's computing architecture to satisfy the future generation needs of diversified technologies and techniques, which brings extended support to the ubiquitous environment. Primitively the industrial 4.0 having a lots of chained interlinked process which also holds valuable information. So it is especially designed for fault tolerance data processing integrated system. This implementation model constructed in the way that smart control and self-accessible system for next generation cyber physical machine and automation controlling system. Also that focusing towards dynamic customization, reusability, eco friendliness for next generation controlling and computation power.

## 1. Introduction

The Industries around the world facing tremendous pressure to delivering their product matching with their competitive brands. For achieving those targets they are putting a lots of efforts on improvising their Methods and technology. In this dynamic environment they are in need to monitoring and controlling those thing in various manner. Finally they are placing various factory to identify and dig those quality and profit from various analytic methods. which is strengthen by the things called integration and finding of cyber physical system leads of creating Bigdata to find the values of business from multiple dimension. And providing the platform for industrial informatics. For supporting those we are adopting various technical and technologies with revolutionary methods listed below.

### 1.1 Industry 4.0

Industry 4.0 revolution is the most comprehensively focusing towards cyber-physical Architecture by which achieving, components like sensors should be provide feature to sense behaviour like Self- Aware and Self-Predictive which leads to degradation monitoring and provide life prediction, which should lead us to production efficiency. Machine controller should aware, predict and compare which leads to maximum up time and predictive health monitoring. In that same aspect production system such as networked manufacturing system provide worry free productivity with its attributes self-configuration, Self-maintain, self-organize. Industry 4.0 it's in the transformation from a manufacturing to a service business model.

*Author for correspondence

### 1.2 IOT

Internet of Things (IOT) can be identified in the three major dimensions from its orientation things, internet and schematic. Internet it's also known as middle ware of this system. Things are sensors which sense and brings information to the system and schematic or knowledge helps to do manipulation. Those where the key factors behind that IOT. Which enables ubiquitous with interconnection to provide accessibility at anytime, anywhere and in any form. Which is creating a lots of challenges in various dimensions such as embedded communication between to achieve intercommunication with the sensors, actuators, etc. Middle ware facing a challenges to provide platform for bigdata analytics and its relevant issues, high cognition, etc[13,3].

### 1.3 Big Data

Big data is the existing term which attracts society today with its extending features to finding business values[2]. In this competitive Business world the term big-data. Which not only represent the volume of data, apart from various dimensions and nature of data unable to process by the existing system or expensive to process those. Such as velocity, variety, verdict and other aspects.

## 2. Problem Statement

From the past decade the system, which is placed inside the Industrial robots are most commonly processed in the master system. PLCs plays vital role in the controlling part with its static nature in the dynamic procedures. But Industry 4.0 needs a sensible work flow needs aware as well as need to react for sensor information. But the transformation and interconnection between the systems based on various platforms are split with various data base and management techniques as a result we failed to make decision based on the data generated by the sub systems. As a result we are aware about target in terms of success and failures but the industry 4.0 needs seamless quality with efficient consumption. For providing the feature to achieving those we are facing a lots of issue in placing the higher end machine in distributed location make complex decision to integrate and making decision as well as cost of that machine also high. It creates issue in integrating and managing unstructured and scalable data wile finding the solution on it.

Like the system need to be respond for the data as well as need to perform some special task. While the system failed the entire chained process also been failed. And the special hardware need to be placed for data conversion and to establishing the communication with the sensors and communicators. For those we are forced to implement a lots of different mechanism and machines to be placed to make everything complex for easy operation. As a result we are getting unique machines with expensive operations in all the dimension.

## 3. Unique Sense Prototype

Figure 1 is the unique sense prototype is the hybrid combination of Hadoop on ARM architecture. The below methodology is the procedural way to combine those two streams for creating the model mentioned in the Figure 2.

## 4. Methodology

There are two possible way to power up our system. But in this we choose micro USB instead of GIPO for achieving quick stability based on available resource. But in this case of providing power to I/O components we continuously give 2A - 5 V to meet basic power consumption requirement. After that Code named wheezy is the one of the

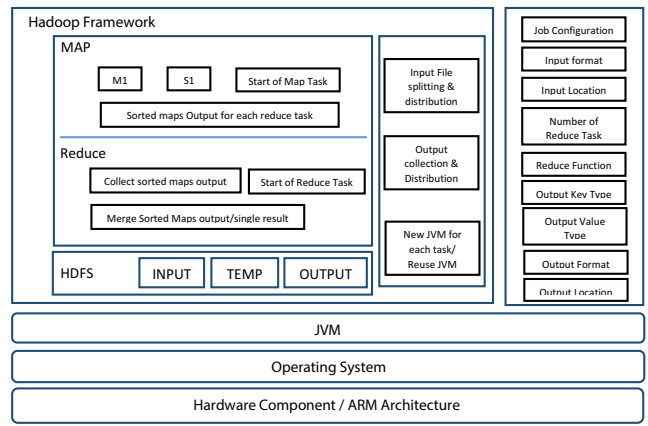

**Figure 1.** Unique Sense: Smart computing prototype[14].

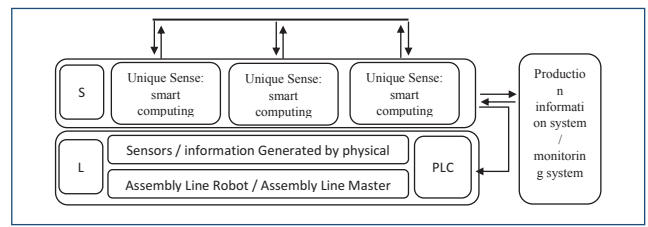

**Figure 2.** Implementation model of IOT in industrial robotic.

S. Vijaykumar, S. G. Saravanakumar and M. Balamurugan

stable version from Debian, Linux distribution. With the future of multi-arch which support 32 bit runs on 64 bit Operating system and its feature extends to support arm[8]. So here in this work we choose it as one of the supporting system for Hadoop on ARM architecture. Therefore, we utilized Rasbian, Debian wheezy linux operating system Kernel Version 3.12, and Released on 9th September 2014 from the Raspberry[9,11] supporting site. Later java has been installed on that architecture because we need JVM for pi. Because Hadoop framework deploy on it to for the execution of separate threads for parallel processing and so on. Here in this prototype we installed openjdk-7 version 1.7.0_07. Then we create new user called hd user especially to avoid collision, later we add it to group to access the file system. Hadoop 1.1.2 is installed on that architecture[1,4] then SSH key created and shared with the required user account for connecting the system remotely using SSH secure shell. Then HDFS created within the linux architecture for that special space allocated with dynamic memory allocation. Then the ownership for accessing that location shared with hd user for file system accessing, with privilege to the user such that 750 is the common type of permission where users can possibly process, read, write and execute (Traverse for directories). It limits the group users for doing the operations only read, execute and denies write operation. It can also avoid data writing violations from other intrusions. hadoop namenode -formate this command formating your file system at the location specified in hdfs-site.xml.

After those installing the required component in Linux, most commonly we need to start the process manually. Here the start-all.sh starts the required components of Hadoop such as name node, data node, secondary name node, job tracker and task tracker. The jps tool lists the instrumented HotSpot Java Virtual Machines (JVMs) on the target system. The tool is limited to reporting information on JVMs for which it has the access permissions[10]. The numeric value represented before the instrumented JVM are its identification number[14].

## 5. Sample Job Result

hduser@raspberrypi /usr/local/hadoop $ hadoop jar hadoop-examples-1.1.2.jar pi 5 50

    Number of Maps  = 5
    Samples per Map = 50
    Wrote input for Map #0
    Wrote input for Map #1
    Wrote input for Map #2
    Wrote input for Map #3
    Wrote input for Map #4
    Starting Job

After the successful job execution we obtain the result with in the duration of 758.142 seconds. Result shown below.

Estimated value of Pi is 3.14800000000000000000

## 6. Model

L – is the Layer which can denote the physical assembly line robots or assembly Line master system or the collection of units, which is capable to process and collect information from that line robot. Which can have sensors and collection of information providing system like data lines such as parallel and sequential cables which can transfer signals and data for processing and retrieving information from the system process in commandment of single PLC or numerous PLCs[7]. Now a day industry 4.0 providing vision to deliver seamless self-aware cyber physical system capable to do work smarter. For that we are focusing various static solutions.

S- Layer are smart computing layer it can be placed in the physical systems in dynamic environment may be a unique or cluster of system based on the industrial need. Which is light weight compact low power consumption model can be work similar like computer but based on the smart ARM architecture[5,6]. Those where well known for 24*7 operation like our mobile phone. From that our unique sense architecture can primarily initiating towards assuring distributed data system and parallel processing across them based on the industrial need such as big data initiation. It can cable to work as individual and work to gather to achieve major tasks. And those where considered in the S layer may have the cluster of networked smart computing units. Those layer can capable to satisfy the IOT properties and efficient for multi operation.

## 7. Discussion

This proposed model is basically an innovative step to next generation industrial findings of industry 4.0 initiation. Now this model is successfully deployed to load data and process those information in parallel manner. So far we facing the issue to collaborate the world of embedded with computation system. But this model can having capable to collect information via GIPO from machineries and establishing the new era of cyber physical system to satisfy the needs of industry 4.0 revolution and it also an IOT, which is capable to support Bigdata Processing, and in addition to that provide platform for data distribution from its hybrid solution. With low power consumption and not necessity to provide any special cooling systems as per the common environment it may vary based on the industrial environment standard.

## 8. Conclusion

The Result proven that the deployment of single node cluster on ARM Architecture successfully executed Pi task in Single board compact portable computer Raspberry-pi. And this system is constructed with in the cost less than 3000 INR. Equivalent to 48$ approximately. This covers the primary unit of the single node architecture excluding the I/O & Displays. This model is capable of collecting sensor information via GIPO and USB which can distribute data among the interconnected Smart computing system with the help of Hadoop framework and also capable to do parallel processing in physically clustered interconnected network architecture with least fault tolerance.

## 9. Future Work

As a continuity of this work, we have planned to introduce multi node cluster on latest ARM Architecture. With Lightweight hard performance and capable to load and distribute data on cost effective model. Introducing Metric for analysing performance and stability of Big data analytics in ARM Architecture along with industrial standard bench marks.